\documentclass[twocolumn,superscriptaddress,showpacs,preprintnumbers,amsmath,amssymb,prb]{revtex4}
\usepackage[dvips]{graphicx}% Include figure files
\usepackage{dcolumn}% Align table columns on decimal point
\usepackage{bm}% bold math
\newcommand{\msr}{$\mu$SR}
\newcommand{\hgro}{Hg$_2$Ru$_2$O$_7$}
\newcommand{\cdro}{Cd$_2$Ru$_2$O$_7$}
\newcommand{\caro}{Ca$_2$Ru$_2$O$_7$}
\newcommand{\tlro}{Tl$_2$Ru$_2$O$_7$}
\newcommand{\hgoo}{Hg$_2$Os$_2$O$_7$}
\newcommand{\cdoo}{Cd$_2$Os$_2$O$_7$}
\newcommand{\lvo}{LiV$_2$O$_4$}
\newcommand{\lzvo}{Li$_{1-x}$Zn$_x$V$_2$O$_4$}
\newcommand{\ymn}{Y$_{1-x}$Sc$_x$Mn$_2$}
\begin{document}

\preprint{APS/123-QED}
\title{Magnetic ground state of pyrochlore oxides close to metal-insulator boundary\\ probed by 
muon spin rotation}% Force line breaks with \\

\author{M. Miyazaki}
\affiliation{Department of Materials Structure Science, The Graduate University for Advanced Studies (Sokendai), Tsukuba, Ibaraki 305-0801, Japan}
\author{R. Kadono}
\affiliation{Department of Materials Structure Science, The Graduate University for Advanced Studies (Sokendai), Tsukuba, Ibaraki 305-0801, Japan}
\affiliation{Muon Science Laboratory and Condensed Matter Research Center, Institute of Materials Structure Science, High Energy Accelerator Research Organization (KEK), Tsukuba, Ibaraki 305-0801, Japan}
\author{K. H. Satoh}
\affiliation{Department of Materials Structure Science, The Graduate University for Advanced Studies (Sokendai), Tsukuba, Ibaraki 305-0801, Japan}
\author{M. Hiraishi} 
\affiliation{Department of Materials Structure Science, The Graduate University for Advanced Studies (Sokendai), Tsukuba, Ibaraki 305-0801, Japan}
\author{S. Takeshita}
\affiliation{Muon Science Laboratory and Condensed Matter Research Center, Institute of Materials Structure Science, High Energy Accelerator Research Organization (KEK), Tsukuba, Ibaraki 305-0801, Japan}
\author{A. Koda}
\affiliation{Department of Materials Structure Science, The Graduate University for Advanced Studies (Sokendai), Tsukuba, Ibaraki 305-0801, Japan}
\affiliation{Muon Science Laboratory and Condensed Matter Research Center, Institute of Materials Structure Science, High Energy Accelerator Research Organization (KEK), Tsukuba, Ibaraki 305-0801, Japan}
\author{A. Yamamoto}
\affiliation{RIKEN, Wako, Saitama 351-0198, Japan}
\author{H. Takagi}
\affiliation{RIKEN, Wako, Saitama 351-0198, Japan}


\begin{abstract}

Magnetism of ruthernium pyrochlore oxides $A_2$Ru$_2$O$_7$ ($A=$ Hg, Cd, Ca), whose electronic properties within a localized ion picture are characterized by non-degenerate $t_{\rm 2g}$ orbitals (Ru$^{5+}$, $4d^3$) and thereby subject to geometrical frustration, has been investigated by muon spin rotation/relaxation (\msr) technique. The $A$ cation (mostly divalent) was varied to examine the effect of covalency (Hg $>$ Cd $>$ Ca) on their electronic property. In a sample with $A=$ Hg that exhibits a clear metal-insulator (MI) transition below $\sim100$ K (which is associated with a weak structural transition), a nearly commensurate magnetic order is observed to develop in accordance with the MI transition. Meanwhile, in the case of $A=$ Cd where the MI transition is suppressed to the level of small anomaly in the resistivity, the local field distribution probed by muon indicates emergence of a certain magnetic inhomogeneity below $\sim30$ K.  Moreover, in Ca$_2$Ru$_2$O$_7$ that remains metallic, we find a highly inhomogeneous local magnetism below $\sim$25 K that comes from randomly oriented Ru moments and thus described as a ``frozen spin liquid" state. The systematic trend of increasing randomness and itinerant character with decreasing covalency suggests close relationship between these two characters.  As a reference for the effect of orbital degeneracy and associated Jahn-Teller instability, we examine a tetravalent ruthernium pyrochlore, Tl$_2$Ru$_2$O$_7$  (Ru$^{4+}$, $4d^4$), where the result of \msr\ indicates a non-magnetic ground state that is consistent with the formation of the Haldane chains suggested by neutron diffraction experiment.  

%, which is in sheer contrast to the response of CuO$_2$ layers against Zn/Ni substitution
\end{abstract}

\maketitle
\section{Introduction}
A class of transition metal oxides with a chemical composition $A_{2}B_{2}$O$_{7}$ crystalizes into so-called cubic pyrochlore structure ($Fd\overline{3}m$),  and it can be regarded as two interpenetrating networks of corner-shared tetrahedra consisting of $A_2$O' and $B_{2}$O$_{6}$ sublattices.\cite{Pannetier:70}  Because of high crystal symmetry, presence of antiferromagnetic correlation or charge disproportion between the metal ions on the pyrochlore lattice induces a geometrical flustration and associated macroscopic degrees of degeneracy for the spin/charge states in their ionic limit.  
This causes a variety of interesting phenomena such as a ``spin glass" phase in $R_{2}$Mo$_{2}$O$_{7}$ with $R=$ Y, Tb and Dy (Refs.\onlinecite{Dunsiger:96,Garner:99}) 
and the ``spin-ice'' phase in $R_{2}$Ti$_{2}$O$_{7}$ with $R=$ Dy and Ho (Refs.\onlinecite{Harris:97,Ramirez:99}).

It is naturally expected that geometrical frustration (or correlation) would also lead to nontrivial effects on itinerant electrons, because they can be a local entity in a certain time scale ($\sim h/\varepsilon_t$, where $\varepsilon_t$ is the relevant energy band width).  The unusual behavior of ordinary and anomalous Hall coefficients in the Mo pyrochlores with $R$=Nd, Sm and Gd might be a manifestation of such effect.\cite{Taguchi:99,Katsufuji:00}
Above all, however, the ``heavy Fermion" behavior  in a metallic vanadium spinel 
LiV$_{2}$O$_{4}$ (in which the V ions form a pyrochlore sublattice) 
is one of the most interesting phenomena among conducting pyrochlores/spinels.\cite{Kondo:97}

In the previous studies on \lvo, we demonstrated by muon spin rotation/relaxation (\msr) and muon Knight shift measurements that, unlike what is anticipated in the conventional 
heavy Fermion metals, local vanadium moments in \lvo\ are not quenched by the Kondo 
mechanism, and that they give rise to a highly inhomogeneous and dynamically fluctuating magnetic ground state that seems to be in close association with the geometrical frustration.\cite{Koda:04,Koda:05}  It would be interesting to note that a strikingly similar situation was reported to occur in an intermetallic Laves phase (C15-type) compound, \ymn, where it is inferred from earlier \msr\ study that inhomogeneous local magnetism persists at low temperatures in a sample that exhibits a heavy Fermion behavior induced by suppression of structural phase transition upon substitution of Y by Sc.\cite{Shiga:93}  Thus, these results demonstrate a pressing need to develop a new concept of quasipartcle state that should represent the anomalous conducting state of electrons under geometrical (three-body) correlation. 

From the experimental view point, it would be helpful for the understanding of anomalous metallic properties to establish the generic character of such electronic states by extending \msr\ study to other electrically conducting pyrochlores/spinels. To this end, spinels and pyrochlores are good candidates for materials survey, although they exhibit a common tendency of structural phase transition at lower temperatures that lifts the orbital degeneracy and relieve the frustration.

\begin{figure}[tp]
\begin{center}
\includegraphics[width=0.5\linewidth,clip]{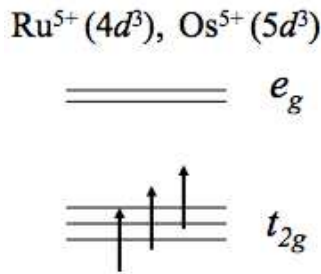}
\caption{
Single ion energy manifold of $B$ site ions
in $A^{2+}_2B^{5+}_2$O$_7$ with $B$=Ru, Os. Three 
$d$-electrons are distributed over different $t_{2g}$ orbitals
($d_{xy}$, $d_{yz}$, $d_{zx}$), so that no orbital
degeneracy is expected.
}
\label{cflevel}
\end{center}
\end{figure}

Here, we focus on a family of ruthenium and osmium pyrochlores, $A_2B_2$O$_7$ ($B$=Ru, Os) with nominally divalent cations (i.e., $A^{2+}_2B^{5+}_2$O$_7$).  As shown in Fig.~\ref{cflevel}, the $B$ site ions have 4$d^3$ (5$d^3$) configuration with the $d$ orbits split into $e_{\rm g}$ doublet and $t_{\rm 2g}$ triplet states (due to the crystal fields of $B$O$_6$ octahedrons), where three electrons are distributed to different $t_{\rm 2g}$ orbits ($xy$, $yz$, $zx$) to satisfy both Pauli exclusion principle and Hund coupling. This means that there is no orbital degeneracy that would lead to the instability against the Jahn-Teller distortion and associated structural phase transition to reduce the crystal symmetry, and therefore one may be able to observe the effect of geometrical frustration at lower temperatures. Moreover, considering that the suppression of structural transition is favorable for preventing the system from MI transition, they would provide a stage for studying the effect of geometrical correlation on itinerant electrons. Interestingly, the actual ground state of these pyrochlores suggested from bulk electronic properties turns out to be considerably dependent on the choice of $A$ site ions. As summarized in Table \ref{pyrolist}, some exhibit metal-insulator (MI) transition that does not necessarily accompany structural phase transition. We have shown by recent \msr\ study that an osmium pyrochlore, \cdoo, which remains in a cubic structure and falls into semi-insulating state below 225 K, exhibits a peculiar magnetic ground state probably related with geometrical frustration.\cite{Koda:07} A similar but more inhomogeneous electronic state is suggested by a preliminary \msr\ study for the case of \hgoo\ which remains metallic down to 4 K (Ref.\onlinecite{Koda:unpub}).

\begin{table}[b]
\begin{center}
\begin{tabular}{cccc}
 & & &\\
\hline\hline
  & SPT ($T_{\rm s}$) & MIT ($T_{\rm MI}$) & $\chi$-drop ($T_{\rm m},T_{\rm g}$)\\
\hline
\hgro &$\simeq107$ K & $\simeq107$ K &  $\simeq107$ K\\
\cdro & No? & No? K  & $\simeq100$ K  \\
\caro & No & No &  SG for $T_{\rm g}\le25$ K\\
\hline
\tlro & $\simeq120$ K & $\simeq120$ K  & $\simeq120$ K \\
\hline
\hgoo & No? & No  & $\simeq80$ K\cite{Koda:unpub} \\
\cdoo  & No? & $\simeq230$ K& $\simeq230$ K\cite{Koda:07}\\
\hline\hline
\end{tabular}
\caption{\small Electronic ground state of Ru and Os pyrochlores, where SPT refers to structural phase transition suggested by X-ray diffraction data, MIT to metal-insulator transition suggested by resistivity data. Magnetic susceptibility ($\chi$) in \hgro\ and \cdro\ exhibits a drop below $T_{\rm MI}$, suggesting development of antiferromagnetic 
spin correlation, whereas a spin glass-like (SG) hysteresis is observed in 
\caro.  (The temperature dependence of $\chi$ are shown in the following sections together with the \msr\ result.)} \label{pyrolist}
\end{center}
\end{table}

In this paper, we report on the magnetic ground state of ruthenium pyrochlore oxides ($A_2$Ru$_2$O$_7$) that are situated near the MI transition. For the nominally pentavalent ruthenides ($4d^3$), the $A$ cation ($\simeq A^{2+}$) is varied among Hg, Cd, and Ca to examine the effect of covalency (Hg $>$ Cd $>$ Ca) on their electronic property. It is shown in samples with $A=$ Hg and Cd that a nearly commensurate magnetic order is observed to develop in accordance with the MI transition (below $\sim$100 K, which is associated with weak structural transition).  However, the local field distribution at the muon site indicates emergence of a certain magnetic inhomogeneity in Cd$_2$Ru$_2$O$_7$ below $\sim$30 K, suggesting that the spin state retains significant amount of entropy at lower temperatures.  Meanwhile in Ca$_2$Ru$_2$O$_7$ that remains metallic, we find a highly inhomogeneous local magnetism below $\sim$25 K that is described by a Gaussian distribution for the local fields at the muon site characteristic those from randomly oriented magnetic moments situated on a regular lattice.  
The observed trends of greater randomness in the magnetism with increasing itinerant character of $d$-electrons is common to that observed in osmium pyrochlores, and thereby it suggests a close link between the degree of magnetic randomness and that of delocalization. As a reference for the effect of orbital degeneracy and associated Jahn-Teller instability, we also examine a nearly tetravalent ruthernium pyrochlore, Tl$_2$Ru$_2$O$_7$  (Ru$^{4+}$, $4d^4$), where the result of \msr\ indicates a non-magnetic ground state that is consistent with the formation of the Haldane chains suggested by nuclear magnetic resonance (NMR)\cite{Sakai:02} and neutron diffraction experiments.\cite{Takeda:99,Lee:06}

\section{Experiment}
Polycrystalline samples of $A_2$Ru$_2$O$_7$ ($A=$Hg, Cd, Ca, and Tl) used in this study were prepared by a solid-state reaction of $A$O and RuO$_2$ together with oxidizer at 950 C$^\circ$ under a high pressure of 4 GPa. The details of sample preparation were described elsewhere.\cite{Yamamoto:07} The bulk properties of these samples were investigated by variety of methods including magnetic susceptibility, electric resistivity, and specific heat measurements. Structural analysis was made using powder X-ray diffraction technique (combined in part with synchrotron radiation). Result of these analyses are summarized in Table \ref{pyrolist}.
In order to obtain sufficient quantity of samples for good signal to noise ratio in \msr\ measuremnts, samples with $A=$Cd, Ca, and Tl were prepared in two batches under the same condition and mixed for exposure to muons after characterization by bulk property measurements.  Their susceptibility data exhibited a slight variance of transition temperatures (i.e., 2--3 K for $A=$Cd and Ca, $\sim5$ K for $A=$Tl).

 Among these compounds, \hgro\ is characterized by a sharp reduction of magnetic susceptibility ($\chi$) below $T_{\rm m}\sim$107 K with a small hysteresis. This occurs in conjunction with increase of resistivity ($\rho$) by nearly one order of magnitude and small lattice expansion accompanying splitting of X-ray diffraction peaks attributed to simple cubic unit cell.\cite{Yamamoto:07} These obserations led to an earlier speculation that spin-singlet ground state might be realized in association with the structural phase transition.  It is interesting to note at this stage that the tetravalent \tlro\ shares many features in common with \hgro: it also exhibits sharp reduction of $\chi$ at $T_{\rm m}\sim 120$ K that is associated with a jump of $\rho$ and structural phase transition.\cite{Takeda:99,Lee:06}  

On the other hand, \cdro\ does not show any clear sign of structural change in the lattice parameters around $T_{\rm m}$, although it exhibits reduction of $\chi$ below $T_{\rm m}\sim$100 K that is similar to \hgro. Resistivity exhibits a rather peculiar temperature dependence, where it is characterized by an anomalous kink around $T_{\rm m}$ and divergent increase at lower temperatures that suggests insulating ground state. Similar features have been reported in an earlier literature.\cite{Wang:98}

Compared with these two compounds, \caro\ is more distinct in the sense that it remains metallic at least above $\sim$2 K; although $\rho(T)$ shows slight enhancement at lower temperatures, it does not obey the relation $\rho(T)\propto\exp(E_a/kT)$ expected for thermally activated carriers over a band gap $E_a$.    More interestingly, $\chi$ exhibits a cusp around $T_{\rm m}\sim25$ K that is typically found in spin glass systems. 
The present sample, which was synthesized by solid-state reaction under high pressure, exhibits these features in common with that previously obtained by hydrothermal method.\cite{Munenaka:06}

Conventional $\mu$SR measurement was performed using the
Lampf spectrometer installed on the M15/M20 beamlines at TRIUMF, Canada. 
 During measurements under zero external field (ZF), residual magnetic field at the sample position was reduced below $10^{-6}$~T while the initial muon spin direction was parallel to the muon beam direction
[$\vec{P}_\mu(0)\parallel \hat{z}$].  For longitudinal field (LF)
measurement, a magnetic field was applied parallel to $\vec{P}_\mu(0)$. Time-dependent muon polarization [$G_z(t)=\hat{z}\cdot \vec{P}_\mu(t)$] was monitored by measuring decay-positron asymmetry along the $\hat{z}$-axis,
\begin{equation}
A(t)=A_0G_z(t)=\frac{N_+(t)-\alpha N_-(t)}{N_+(t)+\alpha N_-(t)},
\end{equation}
where, $A_0$ is the average asymmetry, 
$$N_\pm(t)= N_\pm(0)e^{-t/\tau_\mu}[1\pm A_\pm G_z(t)]$$ 
is the positron event rate for the detector placed in the forward ($+$) or backward ($-$) position relative to the sample, $\tau_\mu$ is the muon decay lifetime ($=2.198\times10^{-6}$ s), $A_\pm$ is the decay positron asymmetry for respective detector ($A_\pm\simeq A_0$), and $\alpha$ is the instrumental asymmetry [$\alpha=N_+(0)/N_-(0)\simeq1$ in usual condition].  Transverse field (TF) condition was realized by
rotating the initial muon polarization so that $\vec{P}_\mu(0)\parallel
\hat{x}$, where the asymmetry was monitored along the $\hat{x}$-axis using signals from an appropriate pair of detectors for $N_\pm(t)$ to obtain $G_x(t)=\hat{x}\cdot \vec{P}_\mu(t)$.  

Because of the limited quantity of samples (0.1--0.2 g) obtained by synthesis under high pressure, a special device (``muon veto counter") was installed to reduce backgrounds due to positron events associated with muons that missed the sample. 
All the measurements under a magnetic field were made by cooling the sample to the target temperature after the field equilibrated.

\section{Result} 

\subsection{\hgro}
\begin{figure}[tb]
\begin{center}
\includegraphics[width=1.0\linewidth]{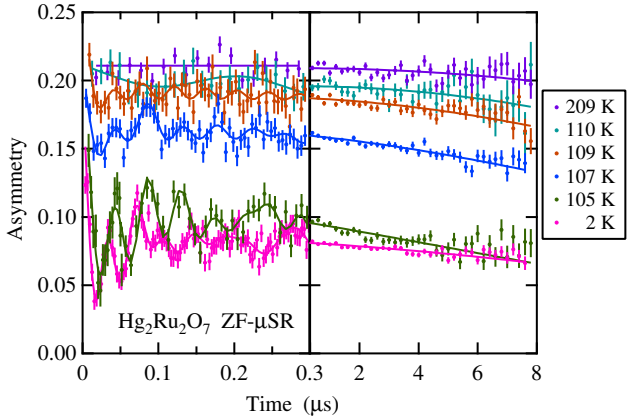}
\end{center}
\caption{(Color online) Time-dependent $\mu$-$e$ decay asymmetry [$A(t)\propto$
muon polarization] observed in \hgro\ at various temperatures under zero external field,  
%Each spectrum is shifted (by 0.1) for  clarity,  
where the full polarization corresponds to $\sim0.21$.  Development of 
spontaneous internal magnetic fields is inferred from the sinusoidal oscillation of 
$A(t)$.
}
\label{spectra_hg}
\end{figure}

\begin{figure}[tb]
\begin{center}
\includegraphics[width=0.8\linewidth]{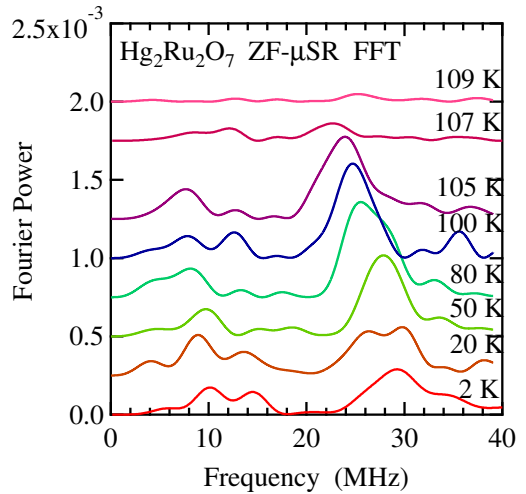}
\end{center}
\caption{(Color online) Fast Fourier transform of the \msr\ spectra shown in Fig.~\ref{spectra_hg}.
Each spectrum is shifted vertically by $2.5\times10^{-4}$ (or $5\times10^{-4}$ between 105 K and 107 K) for clarity.}
\label{fft_hg}
\end{figure}

Fig.~\ref{spectra_hg} shows some examples of ZF-\msr\ time spectra over a temperature range from 2 K to 210 K, where one can observe abrupt emergence of oscillatory signal around $T_{\rm m}\simeq109$ K without enhanced depolarization due to critical slowing down of spin fluctuation in approaching the transition temperature. As is also evident in the Fast Fourier transform (FFT) of these spectra shown in Fig.~\ref{fft_hg}, they indicate step-like development of internal field just below $T_{\rm m}$ associated with a long-range magnetic order.  These features are consistent with suggestion from the hysteresis of $\chi(T)$ that the magnetic transition [which is not obvious solely from $\chi(T)$] is of the first-order, taking place in cooporation with the structural phase transition. 

The FFT spectra are characterized by two bands of dominant frequency lines around 10 MHz and 30 MHz at 2 K, suggesting possibilities of internal field distribution coming from i) a commensurate spin density wave probed at a unique muon site or ii) an antiferromagnetic state with several crystallographically non-equivalent muon sites. Considering these possibilities, we analyzed the time spectra using the following model function valid for the ordered phase of polycrystalline samples,
\begin{eqnarray}
A(t)&\simeq &\sum_{i=1}^nA_i\left[\frac{1}{3}e^{-\lambda_{\rm L}t}+\frac{2}{3}G_z^{\rm KT}(t,\Delta)e^{-\lambda_it}\cos(2\pi f_it+\phi)\right]\nonumber\\
& &\hspace{16em}+A_b,\label{fit}
\end{eqnarray}
which, in the paramagnetic phase, is reduced to
\begin{equation}
A(t)= (A_0-A_b)G_z^{\rm KT}(t,\Delta)+A_b,
\end{equation}
where $G_z^{\rm KT}(t,\Delta)$ is the Kubo-Toyabe relaxation function\cite{Hayano:79} to describe the slow Gaussian depolarization due to random local fields exerted from {\sl nuclear} magnetic moments (with $\Delta\sim10^{-1}$ MHz being the linewidth in the quasistatic limit), 
$A_i$ is the partial asymmetry, $\lambda_{\rm L}$ is the longitudinal relaxation rate (which turned out to be $\simeq0$ for \hgro), $\lambda_i$ is the transverse relaxation rate, $f_i$ is the muon spin precession frequency ($=\frac{1}{2\pi}\gamma_\mu B_i$ with $\gamma_\mu=2\pi\times135.53$ MHz/T and $B_i$ being the local field at the muon site), $\phi$ is the initial phase of precession, and $A_b$ is the background coming from muons that missed the sample ($\sum_iA_i+A_b=A_0$).  When the dynamical fluctuation of local fields is negligible,
we have 
\begin{eqnarray}
G_z^{\rm KT}(t,\Delta)&=&\frac{1}{3}+\frac{2}{3}(1-\Delta^2t^2)\exp(-\frac{1}{2}\Delta^2t^2),\label{ktfunc}\\
&\simeq&\exp(-\Delta^2t^2),\:\:\:(\Delta t\ll1)\nonumber
\end{eqnarray}
for which the original density distribution is given by
\begin{equation}
P_{\rm KT}(B_\alpha,\Delta)=\frac{\gamma_\mu}{\sqrt{2\pi}}
\exp\left[-\frac{\gamma_\mu^2B_\alpha^2}{2\Delta^2}\right]\:\:\:\label{ktdis}
(\alpha=x,y,z).
\end{equation}

For the longitudinal component in Eq.~(\ref{fit}), one must keep in mind that the one third of the implanted muons are subjected to a local field parallel to the initial muon polarization in the magnetically ordered phase of polycrystalline samples, so that the depolarization due to nuclear random local fields are suppressed ($\Delta\ll\gamma_\mu B_i$). We adopted four components ($n=4$) which turned out to be minimal for the satisfactory description of the entire spectra, although the spectra turned out to be dominated by two of these ($i=1$ and 3, see below). 
\begin{figure}[tb]
\begin{center}
\includegraphics[width=0.9\linewidth]{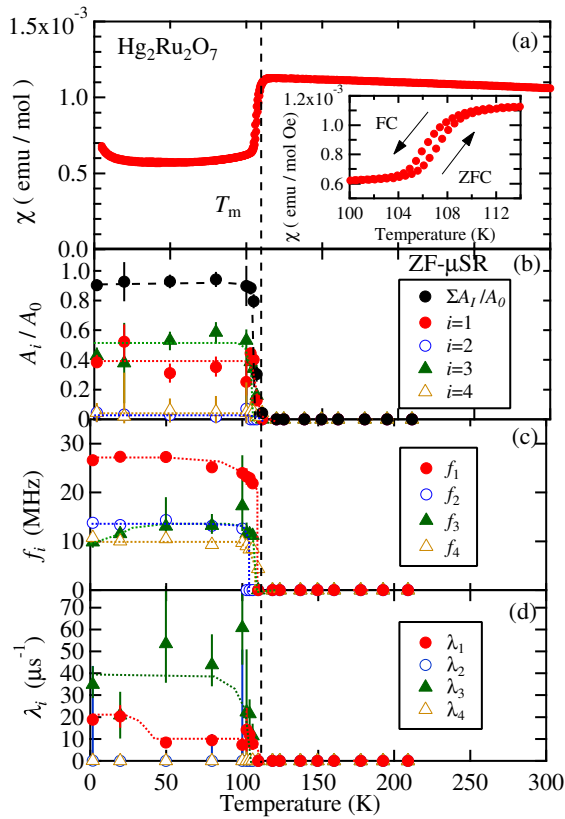}
\end{center}
\caption{(Color online) Temperature dependence of bulk susceptibility (a) and fit parameters obtained by analysis using Eq.~(\ref{fit}), namely (b) fractional yield for the ordered phase defined by $A_i/A_0$, (c) central frequency ($f_i$), and (d) depolarization rate ($\lambda_i$) in \hgro.
}
\label{hg_params}
\end{figure}

As shown in Fig.~\ref{hg_params}, the step-like development of the internal magnetic field below $T_{\rm m}$ is inferred from the temperature dependence of $f_i$ and of the fractional yields for the magnetically ordered phase ($A_i/A_0$). In particular, the total fraction shown in Fig.~\ref{hg_params}(b) indicates that nearly 100\% of the entire crystal falls into fully ordered state just below $T_{\rm m}$: a small background ($A_b/A_0\simeq 5\%$) is attributed to potassium chloride that remains as a residue product of oxidizing agent (KClO$_3$). The depolarization rate of the precession signals is relatively large for the major two components ($\lambda_i\simeq2\pi f_i$).   Moreover, while it is mostly independent of temperature above $\sim50$ K, it exhibits a tendency of increase with decreasing temperature at lower temperatures (see $\lambda_1$) that is also visible in the FFT spectra. As is seen below, this tendency becomes much significant in the case of \cdro.    The small value of the longitudinal relaxation ($\lambda_{\rm L}$) indicates that the local field is static in the time  window of observation ($<10^{-1}$ MHz), and that the large value of $\lambda_i$ is primarily due to the random distribution of local fields at the muon site (not to the dynamical fluctuation).

The Gaussian linewidth, $\Delta$, is determined by local configuration of the  nuclear magnetic moments ($^{99}$Ru, $^{101}$Ru, $^{199}$Hg, and $^{201}$Hg in this case) which are nearest neighboring to muons, and therefore it provides useful information for identfying muon sites. The observed value [$\Delta\simeq0.04(1)$ MHz] has been compared with calculated dipole sum,
\begin{equation}
\Delta =\left[\frac{2}{3}\gamma_\mu^2\sum_j\langle |{\bf \hat{A}}_j{\bm I}_j|\rangle^2\right]^{1/2},\label{ktdlt}
\end{equation}
for several possible muon sites, where ${\bf \hat{A}}_j$ is the dipole tensor,
\begin{equation}
{\bf \hat{A}}_j=\frac1{r_j^3}\left(\frac{3\alpha_i\beta_j}{r_j^2}-\delta_{\alpha\beta}\right)\quad(\alpha, \beta=x,y,z),
\label{dip}
\end{equation}
and the sum is made for the contribution of the $j$-th nuclear magnetic moments ${\bm I}_j$ located at ${\bm r}_j=(x_j,y_j,z_j)$ from a given muon site. As a result, it turned out that an interstitial position near the center of  Ru-cornered tetrahedra is the most probable site (calculated value of $\Delta$ = 0.04 MHz).

Based on this muon site assignment, one can further proceed to examine its consistency with the proposed magnetic structure by comparing the magnetic fields exerted from the Ru$^{5+}$ electronic moments  
\begin{equation}
B_{\rm loc} =|\sum_j{\bf \hat{A}}_j{\bm \mu}_j|
\end{equation}
with the internal field $B_i$ experimentally observed ($\simeq 0.2$ T for $i=1$ and $\sim$0.1 T for $i=$2, 3), where  ${\bm \mu}_j$ is the magnetic moment of Ru ions.  
Recent nuclear magnetic resonance (NMR) study using Hg$^{199}$ and enriched $^{99}$Ru nuclei  suggests a rather complicated magnetic structure consisting of four non-equivalent Ru sites.\cite{Yoshida:10} Considering that i) the internal field is canceled at a quarter of Hg nuclear sites, and that ii) the angle between the primary axis of the electric field gradient and Ru magnetic moment is 90$^\circ$, they suggest a magnetic structure of multiple-${\bf q}$ ($\neq{\bf 0}$) in which Ru spins are antiparallel between those mutually at the diagonal sites of Ru-hexamers on the Kagom\'e lattice (perpendicular to the $[111]$ axis).  Our simulation of internal field at the presumed muon site with further assumption of antiferromagnetic spin structure for triangular lattice layers reproduces the two-band structure of muon precession frequency with the local fields $\sim0.2$ and $\sim0.4$ T/$\mu_B$.  The comparison of these values with the experimental result yields estimated Ru moment size of $\sim0.5(1)\mu_B$, indicating the considerable shrinkage of the moment size form $3\mu_B$.  

%We also note that one of these two frequency bands may correspond to a secondary muon site that is not discernible regarding the magnitude of $\Delta$.

%This, together with the experimental observation of multiplet lines in the ZF-\msr\ spectra suggests a possibility of multiple 
%muon sites. One such possibility would be on the center between two neighboring Ruions.
%While this makes it difficult to examine the consistency on quantitative basis without knowing the muon sites, the complex $\mu$SR line shape is in qualitative agreement with the situation suggested by $^{99}$Ru-NMR.

%We can identify the possible muon site by comparing (i) the observed internal field ($2\pi f_i/\gamma_\mu$) and ${\bm H}_{\rm dip}({\bf r})$ calculated by assuming the orientation and moment size of Cr and Mn ions determined by NPD measurements\cite{Hoshikawa:04}, and (ii) the rms value of the {\sl nuclear} dipolar field, $\Delta/\gamma_\mu\simeq0.35$ mT, observed in the paramagnetic state and that estimated using  eq.~(\ref{dip}) with ${\bm \mu_i}$ substituted by respective nuclear magnetic moments.

\subsection{\cdro}

The ZF-\msr\ spectra in \cdro\ are qualitatively similar to those observed in \hgro. A signal showing spontaneous precession of muon polarization emerges below $T_m\simeq100$ K, indicating that a magnetically ordered state develops below $T_m$. The absence of enhancement in the relaxation rate with decreasing temperature toward $T_m$ is a feature common to \hgro.  It is inferred from the Fourier transform of these spectra that they consist of two bands of frequency lines, which are most obvious around 80 K. However, such a feature, once fully fledged over the intermediate temperature, becomes gradually blurred with decreasing temperature below $\sim50$ K. This is clearly observed by comparing the FFT spectrum at 2 K, where it shows only a  single broad band while the corresponding spectrum  in \hgro\ retains two-band structure (although the latter also exhibits a common tendency of broadening below $\sim50$ K). 
It is also noticeable that the muon precession frequency is considerably reduced from $\sim$15--28 MHz in \hgro\ to $\sim$12--19 MHz in \cdro, suggesting that the effective Ru moment size is accoordingly reduced.

\begin{figure}[tb]
\begin{center}
\includegraphics[width=1.0\linewidth]{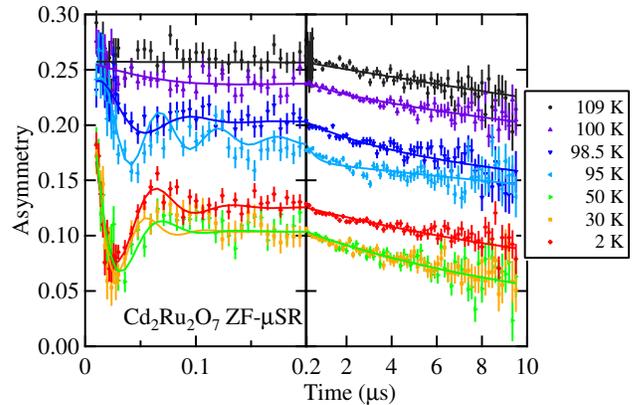}
\end{center}
\caption{(Color online) Time-dependent $\mu$-$e$ decay asymmetry observed in \cdro\ at various temperatures under zero external field.  %Each spectrum is shifted (by 0.1) for  clarity,  where the full polarization corresponds to $\sim0.18$.
}
\label{spectra_cd}
\end{figure}

\begin{figure}[tb]
\begin{center}
\includegraphics[width=0.8\linewidth]{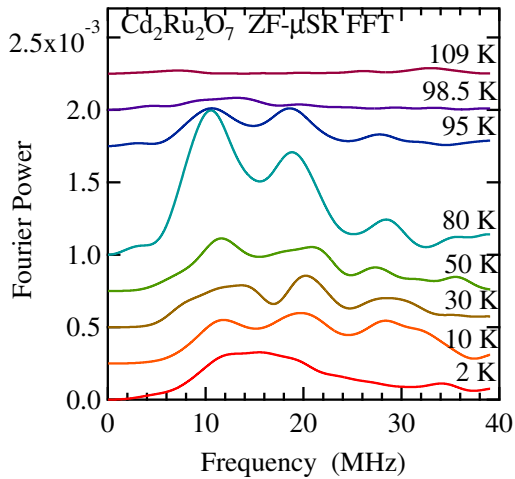}
\end{center}
\caption{(Color online) Fast Fourier transform of the \msr\ spectra shown in Fig.~\ref{spectra_cd}.
Each spectrum is shifted vertically by $2.5\times10^{-4}$ (or $7.5\times10^{-4}$ between 80 K and 95 K) for clarity.}
\label{fft_cd}
\end{figure}

\begin{figure}[tb]
\begin{center}
\includegraphics[width=0.9\linewidth]{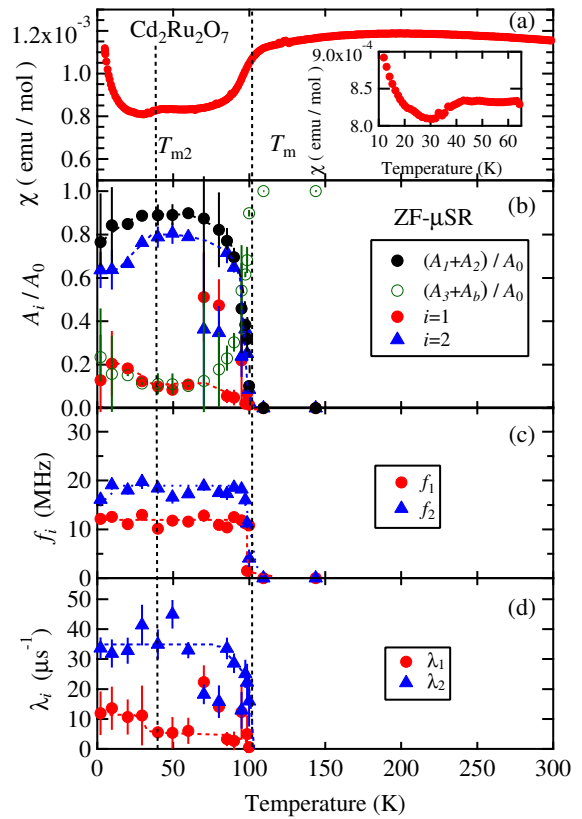}
\end{center}
\caption{(Color online) Temperature dependence of (a) bulk susceptibility (corresponding to one of the two-batch samples) and fit parameters obtained by analysis using Eq.~(\ref{fit}), namely (b) fractional yield for the ordered phase defined by $A_i/A_0$, (c) central frequency ($f_i$, where $f_3=0$), and (d) depolarization rate ($\lambda_i$) in \cdro.
}
\label{cd-params}
\end{figure}

Curve fits of these spectra in time domain using Eq.(\ref{fit}) reveals that we need three components, of which one remains paramagnetic ($f_3=0$) down to 2 K.  The temperature dependence of fit parameters is summarized in Fig.~\ref{cd-params}.  While the internal field develops sharply below $T_m$ (as seen in Fig.~\ref{cd-params}c), the fractional yield of the magnetically ordered state [$(A_1+A_2)/A_0$, Fig.~\ref{cd-params}b] exhibits more gradual increase with decreasing temperature than that observed in \hgro\ (Fig.~\ref{hg_params}d).  Moreover, it exhibits a decrease below $T_{m2}\sim30$--40 K in place of the paramagnetic component ($A_3$). The change is also associated with enhanced depolarization of the first component ($i=1$) whose frequency merges into that of the second component ($i=2$).  It is also noticeable that the bulk susceptibility exhibits a kink near $T_{m2}$ in correspondence with the occurrence of strong depolarization. 

The small relaxation rate for the longitudinal depolarization ($\lambda_{\rm L}$, not shown in Fig.~\ref{cd-params}) at 2 K indicates that the origin of the enhanced depolarization for the precession signal ($\lambda_2$) is mainly due to the enhanced quasistatic randomness of the local fields at the muon site.  This has been also supported by the recovery of the asymptotic component in \msr\ spectra observed under a longitudinal magnetic field (LF).  However, it might be worthy of mention that a broad peak of $\lambda_{\rm L}$ has been observed around 40 K, which suggests slow residual fluctuation of local Ru moments at the intermediate temperature.

It is inferred from the value of $\Delta$ [=0.04283(2) MHz which is close to the case of \hgro] that muon sites in \cdro\ are common to those suggested for the case of \hgro. Then, the observed change of the \msr\ spectra below $T_{m2}$ is ascribed to that in the magnetic structure of Ru moments, which might be associated with the suppression of MI transition in \cdro.  Assuming that the magnetic structure is common to that of \hgro\ for $T_{m2}<T<T_{m}$, the reduction of internal field at the muon site is attributed to the shrinkage of effective Ru moment size to $0.36(7)\mu_B$, which is consistent with enhanced itinerant character of $d$-electrons in \cdro.

\subsection{\caro}

\begin{figure}[tb]
\begin{center}
\includegraphics[width=0.9\linewidth]{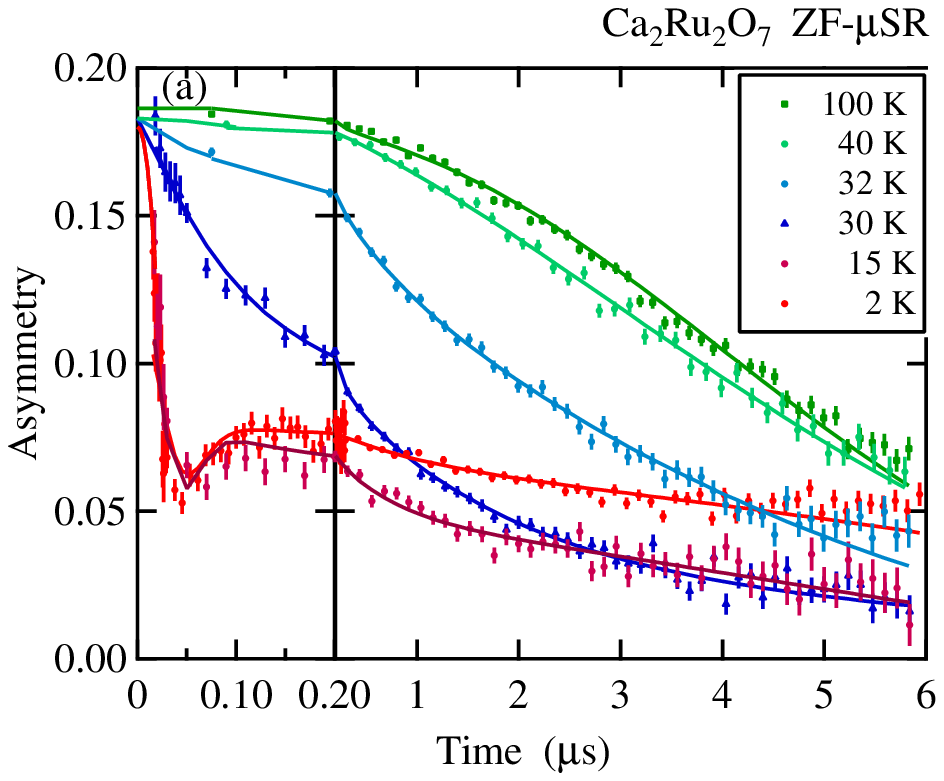}
\includegraphics[width=0.9\linewidth]{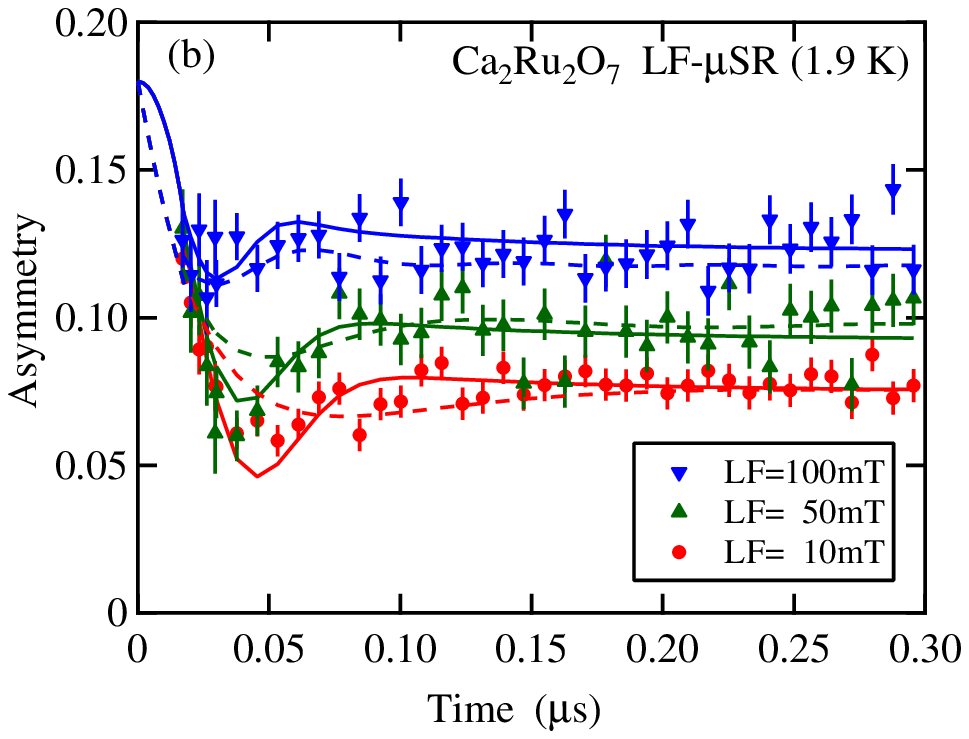}
\end{center}
\caption{(Color online) (a) Time-dependent $\mu$-$e$ decay asymmetry observed in \caro\ at various temperatures under zero external field, and (b) the initial part of the time spectra at 1.9 K 
under a longitudinal field of 10 mT, 50 mT and 100 mT. Solid curves are the best fits using Eq.~(\ref{fit2}) assuming the Gaussian distribution for the local fields (leading to the Kubo-Toyabe relaxation function), whereas dashed curves are those obtained by the Lorentzian distribution (with $\sigma=\Delta_{\rm S}/\sqrt{2}$).  
}
\label{spectra_ca}
\end{figure}

As mentioned earlier, the bulk property of \caro\ is markedly different from the two previous compounds, where the magnetic property inferred from susceptibility is close to that expected for the classical spin glass systems.  The ZF-\msr\ spectra in this compound also exhibits a number of features that are known as those characteristic for the spin glass.  As found in Fig.~\ref{spectra_ca}a, they exhibit increase of depolarization rate and associated change of line shape from Gaussian to exponential-like damping with temperature approaching to the spin glass transition temperature ($T_g\simeq25$ K), indicating the critical slowing down of spin fluctuation towards $T_g$.  However, the detailed analysis reveals that the depolarization is not described by a simple exponential function expected for the case of unique fluctuation rate.  Moreover, the line shape below $T_g$ (which is determined by the distribution of quasistaic local magnetic fields) turns out to exhibit a distinct feature which is not seen in the canonical spin glass like dilute alloys. 

It is well known that the density distribution of internal magnetic field (and associated hyperfine coupling) in the dilute alloys (\underline{Cu}Mn, \underline{Au}Fe, etc., with Mn and Fe concentration less than 3\%) is described by a Lorentzian distribution,
\begin{equation}
P_{\rm L}(B_\alpha,\sigma)=\left(\frac{\gamma_\mu}{\pi}\right)\frac{\sigma}{\sigma^2+\gamma_\mu^2B_\alpha^2}\:\:\:\label{lzdis}
(\alpha=x,y,z)
\end{equation}
with $\sigma$ being the factor determining the width of distribution.\cite{Walstedt:74,Held:75}  This comes from the strong randomness of RKKY interaction which is effectively described by
\begin{equation}
s(r)\sim \frac{\cos(k_{\rm F}r)}{(k_{\rm F}r)^3},
\end{equation}
where $k_{\rm F}$ is the Fermi momentum, and $r$ is the distance between the local magnetic moments.  The distance $r$ varies randomly in the dilute limit while the interaction is long-ranged (decays only by the third power of $r$), leading to freezing of randomly located spins along random orientation. The corresponding time evolution of muon polarization under zero external field is known to have a form
\begin{eqnarray}
G^{\rm L}_z(t,\sigma)&=&\frac{1}{3}+\frac{2}{3}(1-\sigma t)\exp(-\sigma t),\label{lzfunc}\\
&\simeq& \exp(-\frac{4}{3}\sigma t).\:\:\:(\sigma t\ll1)\nonumber
\end{eqnarray}
Apart from the influence of spin fluctuation, the observed \msr\ spectra may be then simply written as 
\begin{equation}
A(t) = A_0G_z^{\rm KT}(t,\Delta)G_z^{\rm L}(t,\sigma)+A_b,\label{fit0}
\end{equation}
which would represent the line shape at the lowest temperature ($T\ll T_g$).
Here, we assume that the hyperfine coupling between the electronic spins of Ru atoms and nuclear spins are small enough so that the spin dynamics of the latter is independent of the former, and that they remain quasistatic within the time window of \msr.  This allows us to make an approximation that the muon depolarization is described by a product of the Kubo-Toyabe function (nuclear-spin part) and exponential decay (electronic part) [which is also implicitly assumed for Eq.~(\ref{fit})]. However,
as shown in Fig.~\ref{spectra_ca}b, the comparison of Eq.~(\ref{fit0}) with the actual spectra observed at 1.9 K indicates that this form does not reproduce the data. More interestingly,
the line shape at 1.9 K is excellently described by the Kubo-Toyabe function with the line width replaced with that for electronic moments, 
\begin{equation}
\Delta_{\rm S} =\left[\frac{2}{3}\gamma_\mu^2\sum_j\langle |{\bf \hat{A}}_j{\bm \mu}_j|\rangle^2\right]^{1/2},\label{deltas}
\end{equation}
where $\Delta_{\rm S}\gg\Delta$ as $|{\bm \mu}_j|\gg|{\bm I}_j|$.  The comparison in Fig.~\ref{spectra_ca}b is made for a set of spectra with different external longitudinal fields ($H_0$) to  enhance the asymptotic component in Eq.~(\ref{lzfunc}) [where these curves can be obtained numerically by replacing $B_z$ in Eq.~(\ref{lzdis}) with  $B_z-\mu_0H_0$, and the same is true for Eqs.~(\ref{ktfunc}) and (\ref{ktdis}) with $\Delta$ replaced by $\Delta_{\rm S}$]. The response of asymptotic component, particularly the dip around $t\simeq0.02$--0.04 $\mu$s and quick recovery to the asymptotic value, indicates that the line shape is described by the quasistatic local fields that obey a Gaussian distribution.

\begin{figure}[tb]

\begin{center}
%\vspace{5cm}
\includegraphics[width=0.9\linewidth]{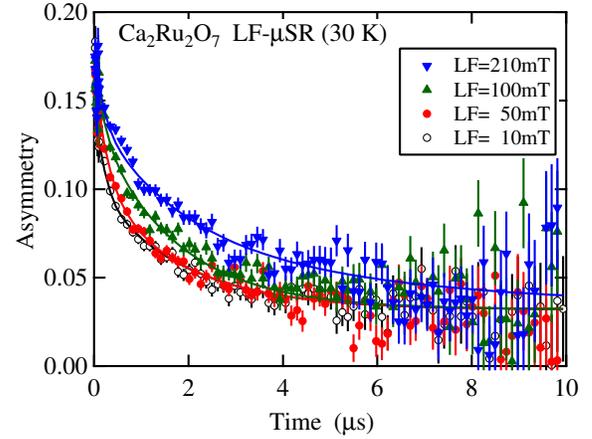}
\end{center}
\caption{(Color online) (a) Time-dependent $\mu$-$e$ decay asymmetry observed in \caro\ at 30 K 
under a longitudinal field of 10 mT, 50 mT, 100 mT and 210 mT. Solid curves are the best fits using Eq.~(\ref{fit2}) assuming three components.  
}
\label{tsplf_ca}
\end{figure}
\begin{figure}[tb]
\begin{center}
%\vspace{5cm}
\includegraphics[width=0.9\linewidth]{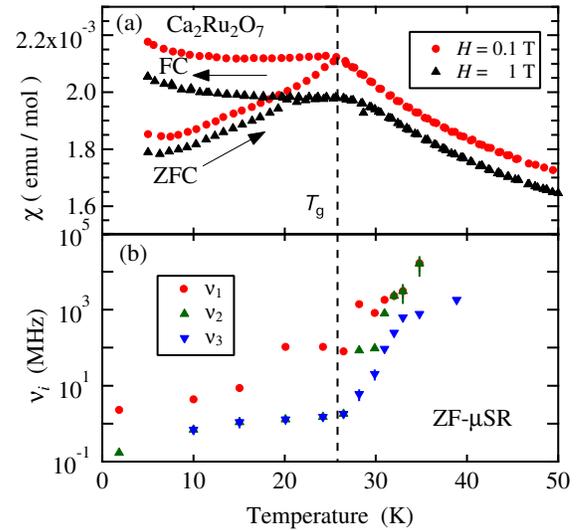}
\end{center}
\caption{(Color online) Temperature dependence of (a) bulk susceptibility  (corresponding to one of the two-batch samples) and (b) Fluctuation rate of the local fields probed by \msr\ in \caro.  Three components with different 
fluctuation rates ($\nu_{1,2,3}$) are presumed in  Eq.~(\ref{fit2}) for $T>T_g\simeq25$ K,  whereas two of them ($\nu_{1,2}$)  for $T<T_g$
(see text for the detail).
}
\label{nu_ca}
\end{figure}

We also find in the detailed analysis that the observed spectra are reasonably reproduced only when we assume multiple components with different fluctuation rates of the local fields, so that Eq.~(\ref{fit0}) should be eventually replaced with
\begin{equation}
A(t)\simeq G_z^{\rm KT}(t,\Delta)\sum_{i=1}^nA_iG_z^{\rm KT}(t,\Delta_{\rm S},\nu_i)+A_b,\label{fit2}
\end{equation}
where $G_z^{\rm KT}(t,\Delta_{\rm S},\nu)$ is obtained numerically by solving an integral equation for the ``strong collision" model\cite{Hayano:79,Kehr:78},
\begin{eqnarray}
G_z^{\rm KT}(t,\Delta_{\rm S},\nu)=G_z^{\rm KT}(t,\Delta_{\rm S})e^{-\nu t}\hspace{6em}\nonumber\\
+\nu\int_0^t e^{-\nu\tau}G_z^{\rm KT}(\tau,\Delta_{\rm S})G_z^{\rm KT}(t-\tau,\Delta_{\rm S},\nu)d\tau,
\end{eqnarray}
which is deduced from the assumption that spin-spin correlation is described by
\begin{equation}
\frac{\hat{s}(0)\cdot \hat{s}(t)}{|\hat{s}(0)|^2}=e^{-\nu t}.\label{sscor}
\end{equation}
The best fits are obtained with two components [$n=2$ in Eq.~(\ref{fit2})] for the spectra below $T_g$  with a Gaussian width $\Delta_{\rm S}=38.5(7)$ MHz (as found in Fig.~\ref{spectra_ca}, the asymptotic component exhibits a slow depolarization corresponding to $\nu_2=2.3(8)\times 10^{8}$ s$^{-1}$ while $\nu_1<10^5$ s$^{-1}$ at 1.9 K), whereas three components ($n=3$) are required to yield reasonable fits for these above $T_g$.  In such a situation, simultaneous fits of multiple spectra obtained under different longitudinal magnetic fields using a common set of parameters provide reliable results.  One such set of data obtained at 30 K yields result of curve fit shown in Fig.~\ref{tsplf_ca}, with $\nu_1=1.48(18)\times10^9$ s$^{-1}$, $\nu_2=1.72(34)\times10^8$ s$^{-1}$, and $\nu_3=0.69(11)\times10^7$ s$^{-1}$ for a common Gaussian width $\Delta_{\rm S}=19.3(5)$ MHz, all having comparable partial asymmetry.  As shown in Fig.~\ref{nu_ca}, these values of spin fluctuation rate spread over three orders of magnitude, strongly suggesting that the spin dynamics of \caro\ cannot be reduced to a simple model like Eq.~(\ref{sscor}) assuming a unique correlation time ($\tau_c\equiv1/\nu$) at a given temperature.

It turns out that the nuclear dipolar width in \caro\ [$\Delta=0.128(2)$ MHz, determined from the spectra for $T\gg T_g$ without much ambiguity] is considerably greater than those observed in \hgro\ and \cdro, suggesting that muon seems to occupy a different site closer to Ru ions: $^{43}$Ca is the only isotope of Ca nuclei with finite nuclear magnetic moments whose natural abundance is negligibly small (0.135\%) and therefore of no relevance to the change of $\Delta$.  However, considering that  both nuclear and atomic dipolar moments are situated in the same Ru site,  the magnitude of $\Delta_{\rm S}$ can be used to evaluate the effective atomic moment size of Ru ions  from that of nuclear magnetic moments using Eqs.~(\ref{ktdlt}) and (\ref{deltas}), yielding 0.69(1)$\mu_B$/Ru ($T<T_g$) or 0.35(1)$\mu_B$/Ru ($T>T_g$).  The latter is in excellent agreement with the value estimated from the temperature dependence of bulk susceptibility 
obtained for single-crystalline specimen of \caro.\cite{Munenaka:06}  

It might be appropriate to point out at this stage that the above mentioned result clearly indicates the different origin of randomness in \caro\ from that in the dilute alloy systems.
This is readily understood by considering the difference in the nature of spin-spin interaction between these two cases. The RKKY interaction in the dilute alloys has a specific feature that its sign changes randomly with relative distance between the local moments located at random. Meanwhile, in the case of pyrochlore compounds, all the magnetic moments are located at the regular lattice position, where the randomness is allowed only for the orientation of the moments: there is no obvious route to introduce randomness 
to the interaction between $B$-site ions. The fact that the local field is described by a Gaussian distribution in \caro\ also indicates the random orientation of Ru$^{5+}$ local moments ($S=3/2$) as the primary source of the randomness, which might be phenomenologically described as a ``frozen spin liquid" state.

\subsection{\tlro}

Unlike other three compounds, the Ru ions in \tlro\ are presumed to be tetravalent (4$d^4$) in the ionic picture, and therefore one of the $4d$ electrons has orbital degeneracy in the $t_{2g}$ manifold.  Such a degenerate state is susceptible to the Jahn-Teller distortion that reduces the free energy by breaking the local lattice symmetry, generally leading to structural phase transition.  As a matter of fact, \tlro\ is known to exhibit a clear structural transition at $T_{\rm s}\simeq 120$ K in conjunction with the metal-insulater (or -semiconductor) transition, where the reduction of symmetry from cubic to orthorhombic structure is inferred from powder neutron diffraction studies.\cite{Takeda:99,Lee:06}  The transition is also associated with a sharp drop of magnetic susceptibility  ($\chi$) at $T_{\rm s}$, where the observed small hysteresis suggests that the reduction of $\chi$ might be driven by the structural transition and associated opening of the band gap.  These bulk properties are quite similar to those observed in \hgro, leading to a naive speculation that the electronic ground state might be common between these two compounds.  

However, it has been suggested from Tl-NMR experiment that the ground state of \tlro\ is non-magnetic, and that it has an excitation gap (as inferred from the behavior of the spin-lattice relaxation rate, $T_1^{-1}$).\cite{Sakai:02}  The recent inelastic neutron scattering experiment also reports occurrence of a well-defined excitation gap ($\simeq11$ meV) below $T_{\rm s}$.\cite{Lee:06}  The gap is attributed to the formation of one dimensional Ru chains ($S=1$) along [110] direction that are understood as a model system of Haldane chains.\cite{Lee:06}  Our \msr\ result supports the presence of non-magnetic ground state, and it provides further evidence for the formation of the Haldane chains in this compound.

\begin{figure}[tb]
\begin{center}
\includegraphics[width=1.0\linewidth]{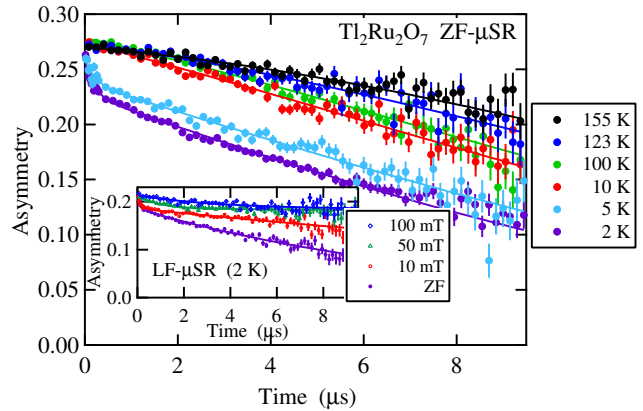}
\end{center}
\caption{(Color online) Time-dependent $\mu$-$e$ decay asymmetry observed in \tlro\ at various temperatures. Each spectrum is vertically shifted by 0.02 for clarity. Inset: the asymmetry observed at 2 K under a longitudinal field of 0, 10 mT, 50 mT, and 100 mT. 
Solid curves are the best fits using Eq.~(\ref{fit3}).  
}
\label{tsp_tl}
\end{figure}

Fig.~\ref{tsp_tl} shows \msr\ time spectra under zero external field observed at various temperatures.
They do not exhibit any anomaly in passing through $T_{\rm s}\simeq 120$ K, where the line shape is mostly like a slow Gaussian depolarization expected for the random local fields from nuclear magnetic moments. (More specifically, the finite gradient $dA(t)/dt$ for $t\rightarrow0$ indicates that an exponential damping due to paramagnetic fluctuation overlaps to the Gaussian damping; see below.)   Thus, the result clearly demonstrates that the new ground state emerging below 
$T_{\rm s}$ does not accompany long-range magnetic order, as has been suggested by Tl-NMR.   Meanwhile, one may notice that the line shape exhibits a gradual change from Gaussian to exponential damping with decreasing temperature, and that a small fraction of the asymmetry shows rapid depolarization below $T^*\sim$10 K.  This quasistatic magnetism can be attributed to  staggered moments at the open edges of Haldane chains,\cite{Hagiwara:90} where the edges may come from various crystalline defects including the twin boundaries inherent in the orthorhombic structure.  Some earlier \msr\ study on another Haldane system (Y$_2$BaNiO$_5$, which happens to have a similar gap energy) indicates that such staggered moments induce spin glass-like random magnetism.\cite{Kojima:95}

\begin{figure}[tb]
\begin{center}
\includegraphics[width=0.9\linewidth]{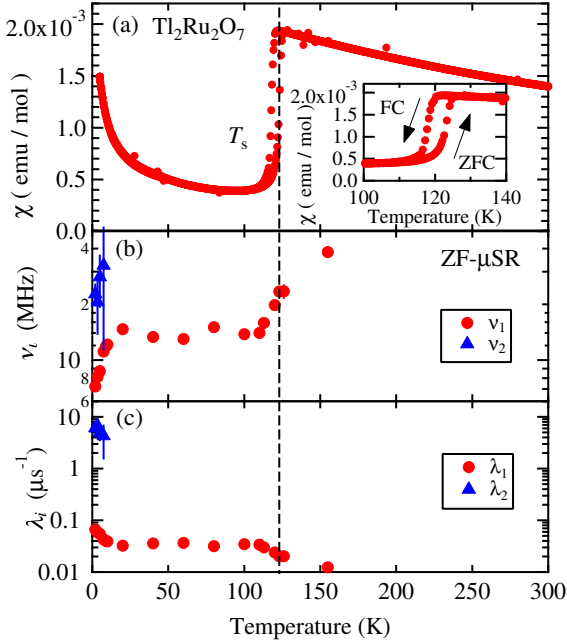}
\end{center}
\caption{(Color online) Temperature dependence of (a) bulk susceptibility  (corresponding to one of the two-batch samples) and fit parameters obtained by analysis using Eq.~(\ref{fit3}), namely (b)  fluctuation rate ($\nu_i$) and (c) corresponding depolarization rate ($\lambda_i$).
}
\label{tl-params}
\end{figure}

Considering these features, we made curve fit analysis of the spectra using a simplified form of Eq.~(\ref{fit0}) with two components, 
\begin{eqnarray}
A(t)& \simeq& G_z^{\rm KT}(t,\Delta)\sum_{i=1}^2A_i\exp(-\lambda_it)+A_b,\label{fit3}\\
 & &\lambda_i=\frac{2\sigma_i^2,\nu_i}{\nu_i^2+\gamma_\mu^2H_0^2},\:\:\:(i=1,2)\nonumber
\end{eqnarray}
which is a good approximation in the paramagnetic state with $\nu_i\ge\sigma_i$. The hyperfine parameters ($\sigma_i$ and $\Delta$) are determined by the simultaneous fit of multiple spectra obtained at 2 K under different longitudinal fields ($H_0$), yielding $\sigma_1=0.49(1)$ MHz, $\sigma_2=8.3(3)$ MHz, and $\Delta=0.042(7)$ MHz.  The value of $\Delta$ indicates that muon site is common to those in \hgro\ and \cdro.  These values are fixed for the rest of analysis to extract the spin fluctuation rates ($\nu_i$) from curve fits for the ZF-\msr\ spectra, where the result is summarized in Fig.~\ref{tl-params}. The second component is observed only below $T^*$
with $A_2$ showing a gradual increase with decreasing temperature ($\simeq 0.03$ at 2 K).

Interestingly, the fluctuation rate exhibits a gradual decrease as temperature is reduced toward $T_{\rm s}$, which is in good correspondence with the behavior of Tl-NMR $T_1^{-1}$ interpreted as a manifestation of reduced spin excitation associated with the development of spin-singlet correlation. 
The fact that it levels off below $T_{\rm s}$ supports the view that \msr\ is also monitoring the same phenomenon through a different time window (i.e., it is not the artifact associated with muon diffusion, for example). 

 Further slowing down of fluctuation ($\nu_1$) is observed below $T^*$, where the second component attributed to the staggered moments appears with intermediate fluctuation rate ($\nu_2$).  It turns out that the emergence of this component occurs in parallel with the appearance of a large Curie term in the magnetic susceptibility.  Thus, the seemingly ``impurity"-like behavior in $\chi$ is explained by the edge states associated with the Haldane chains. 

\section{Discussion}

\subsection{Correlation between magnetic randomness and itinerant character of $4d$ electrons}

We have shown by the \msr\ studies on a series of Ru pyrochlore compounds ($A_2$Ru$_2$O$_7$) that their electronic ground state exhibits a systematic change with varying $A$ cations.  The change in the structural and MI transition temperature suggests that the relevant $4d$ electrons become more itinerant in the system with less covalent $A$ cation (Hg $>$ Cd $>$ Ca).  This is also supported by the effective size of local Ru moments summarized in Table II, which shows tendency of reduction along with the above mentioned order (except below $T_g$ in \caro, where the moment seems restored on the suppression of spin fluctuation).  While it is not clear how the variation in the covalency of $A$O sublattice may affect the  RuO$_6$ network, the $t_{2g}$ band width ($\varepsilon_{d\epsilon}$) seems to be comparable to (or greater than) the energy gain brought by the local lattice distortion. We might stress here that the magnetism in the ground state exhibits stronger randomness in the compounds showing more itinerant character.   The absence of structural change upon MI transition in \cdro\ might even suggest that the transition is of a Mott type ($U\ge\varepsilon_{d\epsilon}$, with $U$ being the electronic correlation energy), as the $t_{2g}$ band is half-filled.  The appearance of more random magnetism at lower temperatures (below $\sim$30 K) suggests that the magnetism just below $T_{\rm m}\simeq 100$ K is not a fully fledged N\'eel state, and that a significant amount of entropy is preserved.  In this regard, it might be worthy of remark that the Dzyaloshinski-Moriya (DM) interaction would serve as a direct driving force for the secondary magnetic phase with an energy scale of 25--30 K (i.e., $\sim k_BT_{m2}$ in \hgro, \cdro, and $\sim k_BT_g$ in \caro\ which are close with each other and suggesting a common origin).  The path of exchange interaction between Ru ions indeed lacks inversion symmetry, and thereby Ru ions are subject to the DM interaction that tends to rotate the Ru moments around the direction of ${\bf S}_i\times{\bf S}_j$. This may be cooperatively serving with the itinerant character of $d$ electrons (see below) to induce the randomness in the electronic ground state of Ru pyrochlores.

\begin{table}[tb]
\begin{center}
\begin{tabular}{ccc}
\hline\hline
  & $T_{\rm m},T_{\rm g}$ & Ru moment size\\
\hline
\hgro &$\simeq107$ K & $0.5(1)\mu_B$ \\
\cdro & $\simeq100$ K & $0.36(7)\mu_B$ \\
\caro & $T_{\rm g}\simeq25$ K & \parbox{10em}{$0.35(1)\mu_B$ ($T>T_g$), $0.69(1)\mu_B$ ($T<T_g$)}\\
\hline\hline
\end{tabular}
\caption{\small Effective Ru moment size in $A_2$Ru$_2$O$_7$ estimated from the magnitudes of local fields at muon site.} \label{moments}
\end{center}
\end{table}

%The complex magnetic structure in \hgro\ suggested by NMR might be a precursor of such a hidden entropy which manifests as a spin glass state in \caro.

According to the two-dimensional $t$-$J$ model on a triangular lattice (in the vicinity of half-filled band, where ``$t$"$\sim\varepsilon_{d\epsilon}$ is the hopping integral, $J$ is the Heisenberg exchange energy)\cite{Koretsune:02,Koretsune:03}, the presence of geometrical frustration leads to the competition in the ground state between the Nagaoka's ferromagnetism (which is rigorously proven for $J=0$ with one hole) and the N\'eel state for $t<0$, leading to the large degeneracy and associated entropy near $T=0$. The strong randomness in the magnetism observed in \cdro\ and \caro\ might be understood as consequence of a similar situation (with decreasing effective $J$ in place of $t$) on the three-dimensional stage, where the quasistatic character of the local spins might imply a small effective $J$ due to the two competing magnetic correlations.

\subsection{Close similarity between \caro\ and \lzvo}  

It has been known in the earlier studies that substitution of Li$^+$ with Zn$^{2+}$ in \lvo\ induces spin glass phase at low temperatures [for $x\ge0.05$ in \lzvo,  where the valence state of the vanadium is V$^{3.5-x/2}$  in the ionic limit].\cite{Ueda:97,Trinkl:00,Krimmel:00}  This observation comprises the experimental basis that the geometrical frustration is at work among vanadium ions which also occupy the pyrochlore sublattice in the spinel structure.  The compound exhibits metallic conductivity except $x>0.9$ where a structural transition (cubic-to-tetragnal) occurs at low temperature ($\simeq50$ K at $x=1$).  Our detailed \msr\ study on \lzvo\ has shown that the fluctuation of vanadium electronic moments ($\ge10^9$ s$^{-1}$) which persists down to the lowest temperature in pure \lvo, is strongly suppressed below 4--5 K for $x=0.05$--0.1, and that a quasistatic random magnetism shows up below $T_g\simeq10$ K for $x\ge0.2$.\cite{Koda:04}  The sensitiveness of V spin fluctuation to the slight variation of band filling suggests that the quater-filled state ($3d^{1.5}$ for $x=0$) in the vanadium $t_{2g}$ band  
may be a ``quantum critical point" (QCP) across the competing ground states.

\begin{figure}[tb]
\begin{center}
\includegraphics[width=0.85\linewidth]{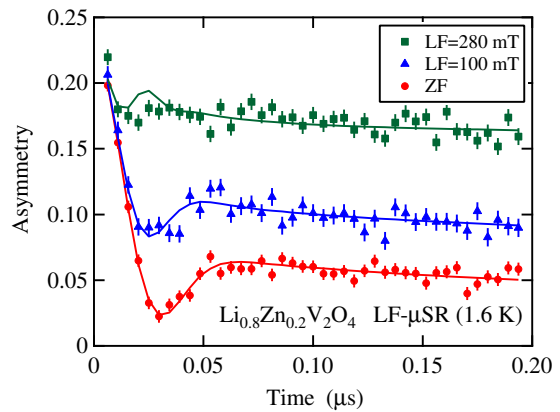}
\end{center}
\caption{(Color online) Time-dependent $\mu$-$e$ decay asymmetry observed in \lzvo\ with $x=0.2$
at 1.6 K under a longitudinal field of 0, 10 mT, 0.1 T, and 0.28 T. 
Solid curves are the best fits using Eq.~(\ref{fit2}) plus a small
fraction of additive exponential damping term (see text for detail).  
}
\label{tsp_lzv}
\end{figure}

Here, we point out that the result of \msr\ in \caro\ shows a marked resemblance with that obtained in \lzvo.  As shown in Fig.~\ref{tsp_lzv}, the line shape of \msr\ spectra observed in the quasistatic spin glass phase of \lzvo\ with $x=0.2$ is well represented by the Kubo-Toyabe relaxation function (where the hyperfine parameter $\Delta_{\rm S}\simeq60$ MHz) with a small fraction ($\sim20$\%) of correction term [an exponential damping added to simulate Eq.~(\ref{lzfunc}) probably due to the randomness induced by Zn substitution for Li], indicating that the density distribution of local magnetic field probed by muon is mostly represented by a Gaussian distribution.\cite{Koda:upd} This again suggests that the distribution is determined by the random orientation of V moments, and that the origin of randomness is common to the case of \caro. 
Considering that the spin glass state is connected to the heavy quasiparticle state in \lvo, the electronic ground state of \caro\ might be also understood as that close to the QCP realized in \lvo.%\vspace{3ex}

\section{Conclusions}

We have shown by muon spin rotation study on a series of ruthenium pyrochlore oxides that the electronic ground state exhibits a systematic variation for different $A$-cations, where the structural and metal-insulator transition temperature ($T_{\rm s}$) is reduced with increasing covalency of the $A$O sublattice.  The reduction of $T_{\rm s}$ and enhanced itinerant character of the compounds is in accordance with the degree of randomness in the ground state magnetism probed by \msr, where the randomness comes from random orientation of the ruthenium local magnetic moments.  Considering a theoretical investigation using the two-dimensional $t$-$J$ model on a triangular lattice, we point out that the origin of such ``spin-orientarion glass" state may come from the competition between the Nagaoka's ferromagnetic correlation and antiferromagnetic exchange interaction in the vicinity of the Mott-insulating  
phase (for the half-filled $T_{2g}$ bands) which is in co-operation with the local Dzyaloshinski-Moriya interaction. 

\begin{acknowledgments}
 We would like to thank the TRIUMF staff for their technical support during the $\mu$SR experiment, and also thank Masao Ogata for helpful discussion.  This work was partially supported by the KEK-MSL Inter-University Program for Oversea Muon Facilities and by a Grant-in-Aid for Creative Scientific Research on Priority Areas from the Ministry of Education, Culture, Sports, Science and Technology, Japan.
\end{acknowledgments}

\end{document}